\documentstyle[12pt,epsf]{article}
\begin{document}
\begin{flushright}{OITS 710}\\
March 2002
\end{flushright}

\vspace*{1cm}

\begin{center}
{\Large {\bf Fluctuation of Gaps in Hadronization
at Phase Transition}}
\vskip .65cm
 {\bf  Rudolph C. Hwa$^1$ and Qing-hui Zhang$^2$}
\vskip.4cm

\vskip.4cm

{$^1$Institute of Theoretical Science and Department of
Physics\\ University of Oregon, Eugene, OR 97403-5203, USA\\
\bigskip
$^2$Department of Physics, McGill University, Montreal,
Quebec H3A 2T8, Canada}
\end{center}
\vskip.3cm

\begin{abstract}
Event-by-event fluctuations of hadronic patterns in
heavy-ion collisions are studied in search for signatures of
quark-hadron phase transition. Attention is focused on a
narrow strip in the azimuthal angle with small $\Delta y$.
The fluctuations in the gaps between particles are quantified
by simple measures. A scaling exponent $\alpha$ is shown
to  exist around $T_c$. An index $\xi$ is shown to characterize the
critical fluctuation; it is a numerical constant
$\xi=0.05\pm0.01$. All the measures considered in this gap
analysis are experimentally observable. Whether or not the
theoretical  predictions, based on simulations using
2-dimensional Ising model, are realistic for heavy-ion
collisions, analysis of the experimental data suggested
here should be carried out, since the existence of a scaling
behavior is of interest in its own right.

\end{abstract}
\newpage
\section{Introduction}

If a quark-gluon system created in a heavy-ion collision at
high energy undergoes a second-order phase transition to
hadrons, one should expect large fluctuations in the
hadron densities from region to region.  Thus in any small
interval of hadronization time the particles produced on
the surface of the system, where the temperature $T$ is
 at the critical temperature $T_c$, should form patterns of
clusters and voids of all sizes.  The question is how such
fluctuations can be detected.  Since the detector
integrates over the hadronization time, the hadron
patterns formed at different times, when added, tend toward
a uniform sum.  In Refs.\ \cite{hz,hz2} we have studied the
problem and found a way to overcome the difficulty.  On
the one hand, one must make severe cuts in the transverse
momentum $p_T$ (accepting particles only in a narrow
window of width $\Delta p_T$), while on the other hand,
appropriate measures that are insensitive to the
randomization process following hadronization should be
used to quantify the critical fluctuations.  We proposed a
void analysis that identifies the scaling properties of the
fluctuations.  The suggested measures in the analysis are in
the two-dimensional space of rapidity $(y)$ and azimuthal
angle $(\phi)$.  The analysis has been applied to the NA49
data of $Pb$-$Pb$ collisions at CERN-SPS \cite{cy}, but
remains to be considered for the data collected at
BNL-RHIC.

In this paper we aim at simplifying the experimental
problem of detecting the critical fluctuation, while making
one of the theoretical parameters in the void analysis more
experimentally relevant.  They are accomplished by
reducing the 2D space to 1D so that a void becomes a gap.
The study of gaps in rapidity was undertaken by us earlier
in connection with multiparticle production in hadronic
collisions \cite{hz3}.  The proposed gap analysis has been
applied to the NA22 data \cite{lw} and provides a way of
testing the predictions of the dynamical models on soft
processes.  We now apply the gap analysis to the phase
transition problem in heavy-ion collisions.  Besides
showing what can be expected theoretically, we believe that
the suggestions made here are more amenable to the
experimental situation.  An important outcome of this
study is the identification of a numerical constant that
characterizes the critical fluctuation.  It can be checked by
experiment.

Our theoretical tool, as in \cite{hz,hz2}, is still the
simulation of critical phenomenon on the 2D Ising lattice.
Not only is the simulation simple to execute, it is also
physically relevant, since the quark-hadron phase
transition is very likely to be in the same universality class
as the Ising system \cite{ggp}.  We map the 2D Ising lattice
to the surface of the quark-gluon plasma cylinder formed in
a central collisions of heavy ions.  We select a row in the
lattice and map that to a strip on the plasma cylinder
confined to a narrow interval in rapidity.  Our proposed
analysis is for the experimental examination of the
fluctuation of gaps between particles in the azimuthal
$\phi$ variable.

\section{Gap Analysis on the Lattice and for the
Experiments}

The relationship between the Ising problem and the hadron
density problem has been discussed in detail in Refs.\
\cite{hz,hz2}.  We summarize here only the essentials for
what is needed in this paper.

We start with a 2D lattice of size $L^2$, where $L = 288$.
On the lattice we define cells of size $\epsilon^2$, where
$\epsilon = 4$.  The density $\rho_i$ at the {\it i}th cell is
defined to be
\begin{eqnarray}
\rho_i =  \lambda c^2_i \theta (c_i)\quad ,
\label{2.1}
\end{eqnarray}
where $c_i$ is the net spin at cell $i$ defined to be
positive along the overall magnetization, i.e.,
\begin{eqnarray}
c_i = \mbox{sgn} \left(\sum_{j\epsilon L^2} \sigma
_j\right) \sum_{j\epsilon i} \sigma_j ,
\label{2.2}
\end{eqnarray}
$\sigma_j$ being the lattice spin at site $j$, and
$\lambda$ an unspecified factor relating $c^2_i$ to the
particle density in $y$-$\phi$ space.  Near $T_c$,
$\rho_i$ fluctuates from cell to cell; however, the existence
of correlation of all length scales leads to clusters of all sizes
formed out of neighboring cells having non-vanishing
$\rho_i$.

In Refs.\ \cite{hz,hz2} we considered bins of size $\delta
^2$, with $\delta$ being integral multiples of $\epsilon$.
We regarded a bin to be empty if the average bin density,
$\bar{\rho}_b$, is less than a threshold density $\rho
_0$.  We then defined a void to be a collection of
contiguous empty bins, connected by at least one side
between neighboring empty bins.  The measure used to quantify
the fluctuation of void sizes was shown to be insensitive to
the value of $\rho _0$.

Without undercutting the importance of doing the void
analysis on experimental data with high statistics,
especially at high energies, we want now to consider in this
paper the possibility of doing a similar analysis in 1D,
since if clustering of hadrons occurs in 2D, they should
also be visible in 1D.  Not only will the statistics improve,
there will also be no need to do binning to define voids, as
we shall see.

Let us consider a row of $C$ cells, where we initially take
$C = L/\epsilon = 72$.  We map that row to a slice of the
plasma cylinder in the central region with width
$\Delta y$ in rapidity.  What $\Delta y$ is will be
discussed below.  We also will limit $p_T$ to a narrow
interval $\Delta p_T$, as discussed extensively in Refs.\
\cite{hz,hz2}.  Thus we have only the azimuthal angle
$\phi$ as our space of variable.  The row of $C$ cells is
mapped to the interval $0 \leq \phi < 2 \pi$.

We simulate $N_e$ configurations (where $N_e = 3
\times 10 ^5$) on the Ising lattice in 2D at $T = T_c$, and
determine the density $\rho _i$ at each cell in each
configuration.  The simulation cannot be done in 1D
because a 1D Ising system does not exhibit critical
behavior.  From a 2D configuration we choose an arbitrary
row and examine the values of $\rho _i$ in each cell with $1
\leq i \leq C$.  The value of $\lambda$ in Eq.\ (\ref{2.1})
is effectively set to 1 in what we do with $\rho _i$, since
we always compare it to a threshold $\rho _0$ in units of
$\lambda$.  We define an occupied cell to be one in which
\begin{eqnarray}
\rho _i > \rho_0  \quad ,
\label{2.3}
\end{eqnarray}
and we place one (and only one) particle at cell $i$.  If
$\rho _i \leq \rho _0$, then it is an empty cell.  In
mapping to the $\phi$ variable, we place a particle at
$\phi _i$ randomly in the interval $(i - 1)2\pi/C \leq \phi
< i2\pi/C $, when the {\it i}th cell satisfies Eq.\ (\ref{2.3}).
Thus after the whole row is mapped to the $\phi$ variable,
there are $N$ particles in the $\phi$ space, where $N$
fluctuates from event to event.  An event corresponds to a
configuration on the lattice.

On the lattice the average number of occupied cells
$\left<N\right>$ in a row, after averaging over all
configurations, depends on the threshold $\rho _0$.  In the
experiment the average number of particles
$\left<N\right>$ in the
$\phi$ space depends on $\Delta y \Delta p_T$.  In our
analysis we shall vary $\rho _0$.  The predicted
dependence of our measure on $\rho _0$ is to be checked
by experiment by varying $\Delta y \Delta p_T$.  In that
sense the variable
$\rho _0$ is not a theoretical parameter devoid of
experimental relevance.

In an experiment of heavy-ion collisions the single-particle
inclusive distribution in $\phi$, $dn/d\phi$, may not be
uniform, because the impact parameter is not always
exactly zero (even if a centrality cut is made), or because
the detector efficiency is not symmetric in $\phi$, mostly
likely both.  It is then better to use the cumulative variable
$X$ \cite{bg, wo}, defined in our case by
\begin{eqnarray}
X(\phi) = \int^{\phi}_0 {dn \over
d\phi^{\prime}} d\phi^{\prime} \left/ \int^{2\pi}_0 {dn
\over d\phi^{\prime}} d\phi^{\prime}\right. .
\label{2.4}
\end{eqnarray}
The range $0 \leq \phi \leq 2\pi$ is thus mapped to $0
\leq X \leq 1$, and the density of particles in $X$,
$dn/dX$, is uniform.  The distribution $dn/d\phi$ is
determined after many events, but the variable $X$ is used
for exclusive distributions event-by-event.  In a simulation
$dn/d\phi$ is uniform, if $N_e$ is large; we nevertheless
use the $X$ variable between 0 and 1 just so that our
proposed gap analysis corresponds to what is to be done
with experimental data.

Consider an event with $N$ particles in our window,
labeled by $i = 1, \cdots, N$, located in the $X$ space at
$X_i$, ordered in accordance to $X_i < X_{i + 1}$.  Now
define the distance between neighboring particles by
\begin{eqnarray}
x_i = X_{i + 1} - X_i, \qquad \qquad i = 0, \cdots, N ,
\label{2.5}
\end{eqnarray}
with $X_0 = 0$ and $X_{N +1} = 1$ being the boundaries
of the $X$ space.  We call these $x_i$'s gaps, which clearly
satisfy
\begin{eqnarray}
\sum^N_{i = 0} x_i = 1 .
\label{2.6}
\end{eqnarray}
Let us define for each event the moments
\begin{eqnarray}
G_q = {1 \over N + 1} \sum^N_{i = 0}  x_i^q  ,\label{2.7}
\end{eqnarray}
which, for positive $q$, puts more emphasis on the large
gaps than the small ones.  The set $\{G_q\}$ with $2 \leq
q \leq Q$, $Q$ being some number less than 10, say, will
be our quantification of the event structure.  The set
fluctuates from event to event, especially at the critical
point.  Our method is to use $\{G_q\}$ as a basis to
construct a measure that can reveal the critical behavior.

Among the various erraticity measures considered in Ref.\
\cite{hz3} we choose $S_q$ for our consideration here,
since it has the simplest behavior.  Starting with the
definition
\begin{eqnarray}
s_q = \left<G_q \ln G_q\right> ,
\label{2.8}
\end{eqnarray}
where the average $\left<\cdots\right>$ is performed over
all events,
$S_q$ is defined by
\begin{eqnarray}
S_q = s_q / s_q^{st}
\label{2.9}
\end{eqnarray}
where $s_q^{st}$ is as defined in Eq.\ (\ref{2.8}) but with
the statistical contribution to $G_q$ only.  Thus the
deviation of $S_q$ from 1 is a measure of how strong the
dynamical fluctuations are relative to the statistical
fluctuations.  In our simulation we generate $N$ random
numbers between 0 and 1 to populate the $X$ space for the
calculation of
$G_q^{st}$.  In an experiment one can do something
similar to generate a random event.

We shall study the dependence of $S_q$ on $q$ for various
values of $\rho _0$, using configurations simulated on the
Ising lattice.  The same can be done with the experimental
data from heavy-ion collisions.  If hadrons are formed in
the latter due to a second-order phase transition, then the
experimental $S_q$ should reveal some features that are
similar to our theoretical $S_q$ to be presented below.

\section{Results on Critical Fluctuations}
Using $N_e = 3 \times 10^5$ configurations simulated on
the Ising lattice at $T = T_c = 2.315$ in units of $J/k_B$ \cite{hz},
where $J$ is the near neighbor coupling and $k_B$ the
Boltzmann constant, we have determined
$S_q$ for a range of $\rho _0$.  For $\rho _0 = 20$ the
dependence of $S_q$ on $q$ is shown in a log-log plot in
Fig.\ 1.  The solid line is a straight-line fit of the calculated
result, giving strong evidence for the power-law behavior
\begin{eqnarray}
S_q \propto q^{\alpha} .
\label{3.1}
\end{eqnarray}
The value of the exponent is $\alpha = 1.67 \pm 0.02$.
The fact that $\ln S_q$ deviates unambiguously from 0
implies that
$S_q$ is a statistically significant measure of nontrivial
dynamical fluctuation.  The power-law behavior is not
necessarily a consequence of the dynamics of critical
phenomenon, since similar behavior has been found
before in the case of hadronic collisions \cite{hz3}.
However, the exponent $\alpha$ is indicative of critical
fluctuations; it is an order of magnitude larger than  the
value of
$\alpha = 0.156$ obtained for $pp$ collisions at
$\sqrt{s}=20$ GeV.

To have one exponent $\alpha$ for all $q$ is a very
economical description of the critical fluctuations. We can
then investigate the dependence of $\alpha$ on $\rho_0$.
Before so doing, let us first find a replacement of $\rho_0$
by some quantity that is directly measurable. In the
preceding section we have related $\rho_0$ to the
intervals $\Delta y
\Delta p_T$ as the variable under experimental control that
can be used to tune the average multiplicity
$\left<N\right>$ accepted in the $\phi$ window. Thus
$\left<N\right>$ is a quantity that is both experimentally
observable and theoretically computable on the lattice. It is
therefore a suitable replacement for
$\rho_0$. Nevertheless, we prefer to use an even better one
that is the average number of gaps $\left<M\right>$,
which is also observable. On the lattice in accordance to our
convention of counting gaps in Eqs. (\ref{2.5})  and
(\ref{2.6}), we have simply $M = N + 1$.  However, it
should be recalled that in our simulation we have adopted
the rule that, when
$\rho_i$ exceeds $\rho_0$, only one particle is placed in a
cell, not more, no matter how high $\rho_i$ is. That
procedure makes sense in view of our measure $G_q$ of
the event structure, since a tightly packed cell with many
particles in it would have very small gaps that make
negligible contribution in Eq. (\ref{2.7}). Experimentally, if
there are particles whose momenta are indistinguishable, or
nearly so, whether they are separately counted or not also
makes no significant difference in the calculation of
$G_q$.  Thus to allow for such possibilities it is better to
count the number of distinguishable gaps, rather than the
number of particles.

On the lattice we have simulated the configurations for a
range of $\rho_0$ from 20 to 200. For every value of
$\rho_0$ we can calculate the average number of gaps,
$\left<M\right>$. Figure 2 shows the dependence of
$\left<M\right>$ on $\rho_0$ at
$T_c$. The straight-line segments are just the linear
interpolations between neighboring points of $\rho_0$
where the calculation has been made. With such a definitive
relationship at hand, any quantity that depends on
$\rho_0$ will in the following be shown as a function of
$\left<M\right>$, so as to render it amenable to
experimental verification.

For each of the higher values of $\rho_0$ examined, we
have found power-law behavior of $S_q$, as in Fig.\ 1.
Thus the exponent $\alpha$ can be determined in each
case. In Fig. 3 we show the dependence of $\alpha$ on
$\left<M\right>$. Remarkably, the dependence is
very linear. If we parameterize it as
\begin{eqnarray}
\alpha = \alpha_0 + \xi  \left<M\right> ,
\label{3.2}
\end{eqnarray}
we obtain
\begin{eqnarray}
\alpha_0 = -0.258, \qquad\qquad \xi = 0.055 .
\label{3.3}
\end{eqnarray}
The nature of critical fluctuation is
now seen to be reduced to a simple formula, Eq.\
(\ref{3.2}), when the moments of gaps are used to describe
the event structure. If we further put emphasis on the
property that is independent of
$\left<M\right>$, then the slope
$\xi$ in Eq.\ (\ref{3.3}) emerges as a numerical output of
the theory that relies on no numerical input. This is perhaps
the most succinct characterization of the critical
phenomenon, beside the critical exponents. The latter
depend on the temperature of a critical system near $T_c$.
In heavy-ion collisions $T$ is not directly observable, so the
corresponding critical exponents cannot be measured
experimentally (see, however, Ref.\ \cite{hw}). Here we
have an index $\xi$ that is eminently measurable and is
the only numerical constant that can be meaningfully
associated with critical fluctuation.

So far our study has been done only at $T=T_c$. To see how the
results change when $T$ deviates from $T_c$, we repeat the
above analysis for a range of $T$. Figure 4 shows $S_q$ vs $q$ in
log-log plot for $2.27<T<2.80$ in units of $J/k_B$. Note that
linearity is quickly lost when $T$ goes below $T_c=2.315$. For
$T>T_c$ the linearity persists for a limited range $2.315<T<3.1$,
but the slope is reduced. For $T>3.2$ the dependence bends over
at high $q$ and the linearity is lost. For the range of $T$ where
$\alpha$ can be determined from linear fits, we show  in Fig.\
5 $\alpha(T)$ up to $T=2.8$.  The sharp peak that occurs at
$T=T_c$  is a remarkable manifestation of the critical
behavior. For $T$ less than $T_c$, so many hadrons are produced
that the gap distribution rapidly tends toward the statistical. For
$T$ larger than $T_c$, fewer hadrons are produced, and it takes
more $T-T_c$ difference for the statistical fluctuation to
dominate. If
$T$ were experimentally controllable, the measurement of
$\alpha(T)$ would be an excellent way to determine the critical
temperature of the quark-gluon system. That being not the case
in reality, we can only learn from Figs.\ 4 and 5 that, if the
system hadronizes in a range of $T$ around $T_c$, the most
significant portion of the contribution to $S_q$ would come from
the immediate neighborhood of $T=T_c$, and that only the
$\alpha$ value around $T_c$ is experimentally relevant.

The above analysis is done for $\rho_0=20$. We can, of course,
repeat the analysis for other values of $\rho_0$. In each case we
find the peak value of $\alpha$ at $T_c$. At each $T$ where
$\alpha$ can be meaningfully determined, we can investigate its
dependence on $\left<M\right>$, just as we have done in Fig.\ 3
at $T_c$. In each case a linearity is found that allows us to
determine the slope index $\xi$ in Eq.\ (\ref{3.2}). The result is
shown in Fig.\ 6, where $\xi(T)$ also exhibits a peak at $T_c$. We
can now conclude that if hadronization is to occur around $T_c$,
we expect the measurable value of $\xi$ to be around 0.05.

Having established the properties of the observables as
functions of $T$ due to  the dynamics of the critical system, we
now investigate the question of stability with respect to changes
in the size of the detector window, which is kinematical. We have
so far considered a lattice of size 72x72 cells, from which we
choose one row of $C = 72$ cells from each configuration. We
now want to vary the length and width of the row.  We shall do
so by setting $T$ at $T_c$.  First, we consider a row of $C = 54$
cells from each configuration, and later a row of
$C = 36$ cells. In each case it does not matter whether the row is
mapped to a correspondingly shorter range in $\phi$, since
$dn/d\phi$ is converted to
$dn/dX$ in the $X$ space before the gap analysis is
performed. However, the average number of gaps
$\left<M\right>$ does change, so the simulation with
shorter rows does correspond to a shorter ranges of $\phi$
in the experiment. In Fig.\ 7 we show the result on
$\alpha$ vs.
$\left<M\right>$ for the $C = 72, 54$, and $36$ cells.
The straight lines are linear fits, which are evidently very
good. The values of the slope
$\xi$ are :
\begin{eqnarray}
\xi= 0.055 \pm 0.005,&\qquad \qquad C= 72 \nonumber \\
0.057 \pm 0.005,&\qquad \qquad C=54\nonumber \\
0.048 \pm 0.007,&\qquad \qquad C=36
\label{3.4}
\end{eqnarray}
The case of $C = 36$ cells yields a slightly
lower value and relatively larger errors, as is reasonable,
since there are fewer points in a shorter range.
Nevertheless, all three values of $\xi$ agree within errors.
We therefore conclude that $\xi$ is stable against
variations in the length of the $\phi$ window.

Next, we consider variations in the width of the $\phi$
window. In our simulation so far we have used only rows
with 1-cell width. In an experiment the width $\Delta y$ of
a window in $\phi$ will have to be adjusted in order to vary
$\left<M\right>$. To check the stability of our result against
variations in the width of our strip on the Ising lattice, we
consider combinations of two and three rows. Let $k$
be the row index, and $i$ the cell index in the row, as
before. Thus the location of a cell on the lattice is now
denoted by $ki$. We take $r$ adjacent rows and
combine them by performing vertical average of $r$ cells at
each horizontal position $i$ , i.\ e., we define
\begin{eqnarray}
\rho_i = {1  \over  r} \sum^r_{k=1}\rho_{ki}
\label{3.5}
\end{eqnarray}
and
then proceed in the gap analysis using $\rho_i$ as in the
preceding section. It is important that the $r$ rows be
adjacent so that the short-range vertical correlation of the
cells can influence the value of  $\rho_i$ so as to reflect its
dynamical content; otherwise, if the $r$ rows are randomly
chosen, the averaging in Eq.\ (\ref{3.5}) tend to render
 $\rho_i$ more statistical and its distribution in $i$ more
uniform. In short, taking $r$ adjacent rows corresponds to
widening
$\Delta y$ in the experiment. In our calculation we
consider only the case $r=2$ and $3$. Note that we take
the average in Eq.\ (\ref{3.5}) and then vary $\rho_0$ to
change
$\left<M\right>$. We could have added $\rho_{ki}$
(without dividing by $r$) and not vary $\rho_0$; that is
equivalent to averaging $\rho_{ki}$ and dividing $\rho_0$
by
$r$. Our chosen procedure allows more continous change in
$\rho_0$.

The result of our analysis for $r=1, 2$ and $3$, and
$C = 72$, are shown in Fig.\ 8. Roughly speaking, the slopes
are essentially the same within errors, so $\xi$ may be
regarded as independent of $r$. More precisely, we find
\begin{eqnarray}
&\xi = 0.055 \pm 0.005, \qquad \qquad r= 1,\nonumber\\
&\xi = 0.044 \pm 0.003, \qquad \qquad r= 2,\nonumber\\
&\xi = 0.045 \pm 0.003, \qquad \qquad r= 3.
\label{3.6}
\end{eqnarray}
The slight decrease in $\xi$, as $r$
increases, is not unreasonable, since averaging over $r$
rows tends to suppress dynamical fluctuations. In the limit
$r$ becoming very large, clearly only statistical
fluctuations remain. We may summarize Eqs.\ (\ref{3.4}),
(\ref{3.6}) and Fig.\ 6 by the grand average
\begin{eqnarray}
\xi = &0.05 \pm 0.01.
\label{3.7}
\end{eqnarray}
If this
can be verified by experiments, then one should be able to
claim that a signature of critical transition has been
observed. If the value of $\xi$ is not confirmed, yet the
power-law behavior of Eq.\ (\ref{3.1}) is shown to exist in
the data, with or without the linear dependence of
$\alpha$ on
$\left<M\right>$ in Eq.\ (\ref{3.2}), that would still be an
exciting experimental finding, suggestive of dynamical
fluctuations.

There is the usual question about final-state interaction and
the dilution of the dynamical signature due to
randomization. The answer depends on the type of measure
for that signature. The issue has been addressed in
Ref.\ \cite {hw}, where the dependence on the number of
steps of final-state scatterings is examined, in Ref.\
\cite{hz} where configuration mixing is considered, and in
Ref.\ \cite{hz2} where different options in simulating time
evolution are investigated. What one learns from all those
studies is that the measures considered are not strongly
affected by the final-state randomization. Since the gap
analysis is a derivative of the void analysis \cite{hz, hz2},
the same conclusion follows here. The one reminder that we
should emphasize is that the window $\Delta p_T$ in the
transverse momentum should be kept as small as possible to
minimize the overlap of particles emitted at different times.

\section{Conclusion}

We have studied the problem of even-to-event fluctuations
of the hadronic patterns in phase space in heavy-ion
collisions in search for detectable signatures of second-order
quark-hadron phase transition.  We have reduced the
complication of voids in two  dimensions to the simpler
problem of gaps in one dimension. Using the moments of
gaps to construct an entropy-like measure $S_q$, we have
found a power-law dependence on $q$ with an exponent
$\alpha$. It is the dependence of $\alpha$ on the average
number of gaps that yields the index $\xi$, which serves to
characterize critical fluctuation. We have found the stability
of $\xi=0.05 \pm 0.01$ against variations in the length and
width of the detector window in $\phi$. When the
temperature of the system is moved away from $T_c$, the
power-law behavior of $S_q$ on $q$ persists in a narrow range of
$T$ around $T_c$, and the values
of $\alpha$ and $\xi$ show strong peaks at $T_c$. Thus $\xi$ is a
measure of the critical behavior and is a number that arises out
of the study of fluctuations without any numerical input. It is
highly significant that $\xi$ can be checked by experiments,
since all  measures leading to its determination are
designed to be observable.

In heavy-ion experiments it is not difficult to make cuts in
$\Delta p_T$ and $\Delta y$ to limit the average number
of particles in a narrow strip in $\phi$ to the range from
10 to 40. If  those particles are found to be
non-uniformly distributed for every event, the gap
analysis proposed here is  a way to quantify those fluctuations.
When the exponent $\alpha$ and the
index $\xi$ are found to exist, there are numerous variables
under the control of the experiments to vary. We can
mention, for example, the position of the
$y$ slice, the value of $p_T$, the centrality of collisions, the
nuclei sizes and the c.m. energy. It would be very interesting
to see the dependence of $\xi$ on the total transverse
energy $E_T$, since it can provide us with some idea of when
the critical behavior is lost as the nuclei overlap gets to be
too small to create a quark-gluon plasma. Also, for nonzero
impact parameter, even if critical fluctuations exist, the
index $\xi$ may depend on which sector of the $\phi$ space
the analysis is performed, since there is no $\phi$ invariance
in elliptic flow. One can envision a rich variety of
phenomenological studies once the exponent $\alpha$ and
index $\xi$ are found in the experimental data.  They can
provide valuable information about the quark-gluon system.
The application of the gap analysis to the data is therefore
strongly urged.

\section*{Acknowledgment}

   This work was
supported, in part,  by the U.\ S.\ Department of Energy
under Grant No. DE-FG03-96ER40972, and the Natural
Science and Engineering Research Council of Canada and
the Fonds FCAR of the Quebec Government.

\newpage
\begin{center}
\section*{Figure Captions}
\end{center}
\begin{description}

\item[Fig.\ 1]\quad $S_q$ vs $q$ in log-log plot for $T=T_c$ and
$\rho_0=20$.

\item[Fig.\ 2]\quad The average number of gaps $\left<M\right>$
vs $\rho_0$ at $T_c$.

\item[Fig.\ 3]\quad The exponent $\alpha$ vs $\left<M\right>$ at
$T_c$.

\item[Fig.\ 4]\quad $S_q$ vs $q$ in log-log plot for a range of $T$
at $\rho_0=20.$

\item[Fig.\ 5]\quad $\alpha$ vs $T$ at $\rho_0=20$.

\item[Fig.\ 6]\quad $\xi$ vs $T$ at $\rho_0=20$.

\item[Fig.\ 7]\quad $\alpha$ vs $\left<M\right>$ at $T_c$ for
various number of cells in a row.

\item[Fig.\ 8]\quad $\alpha$ vs $\left<M\right>$ at $T_c$ after
averaging over various number of  rows.

\end{description}

\newpage

\begin{figure}[t]\epsfxsize=14cm \epsfysize=14cm
\centerline{\epsfbox{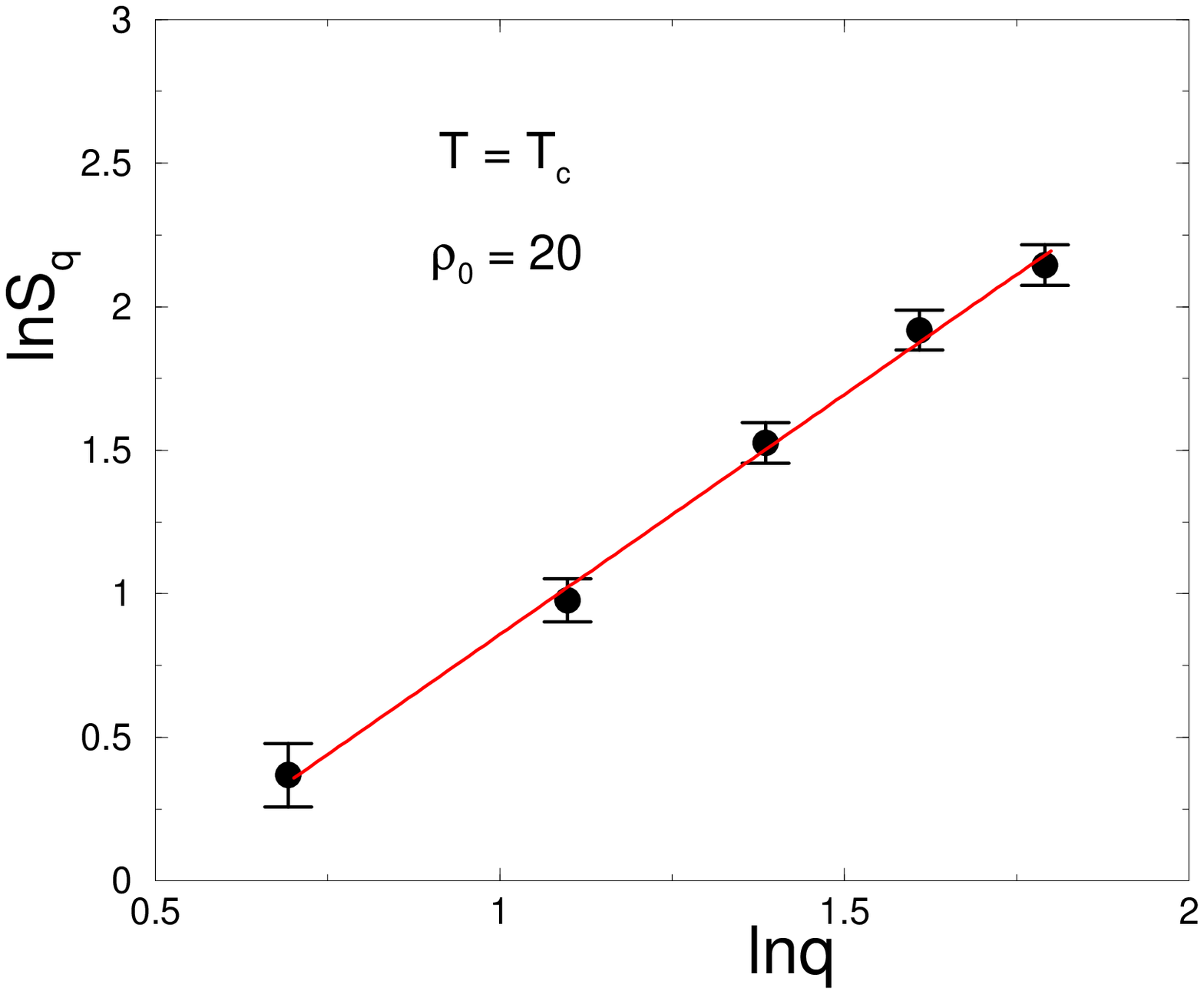}}
\vskip 1.5 cm
\centerline{\bf \large ~~~~~~~~~~~~~~Fig.1}
\vskip -0.5 cm
\end{figure}

\newpage

\begin{figure}[t]\epsfxsize=14cm \epsfysize=14cm
\centerline{\epsfbox{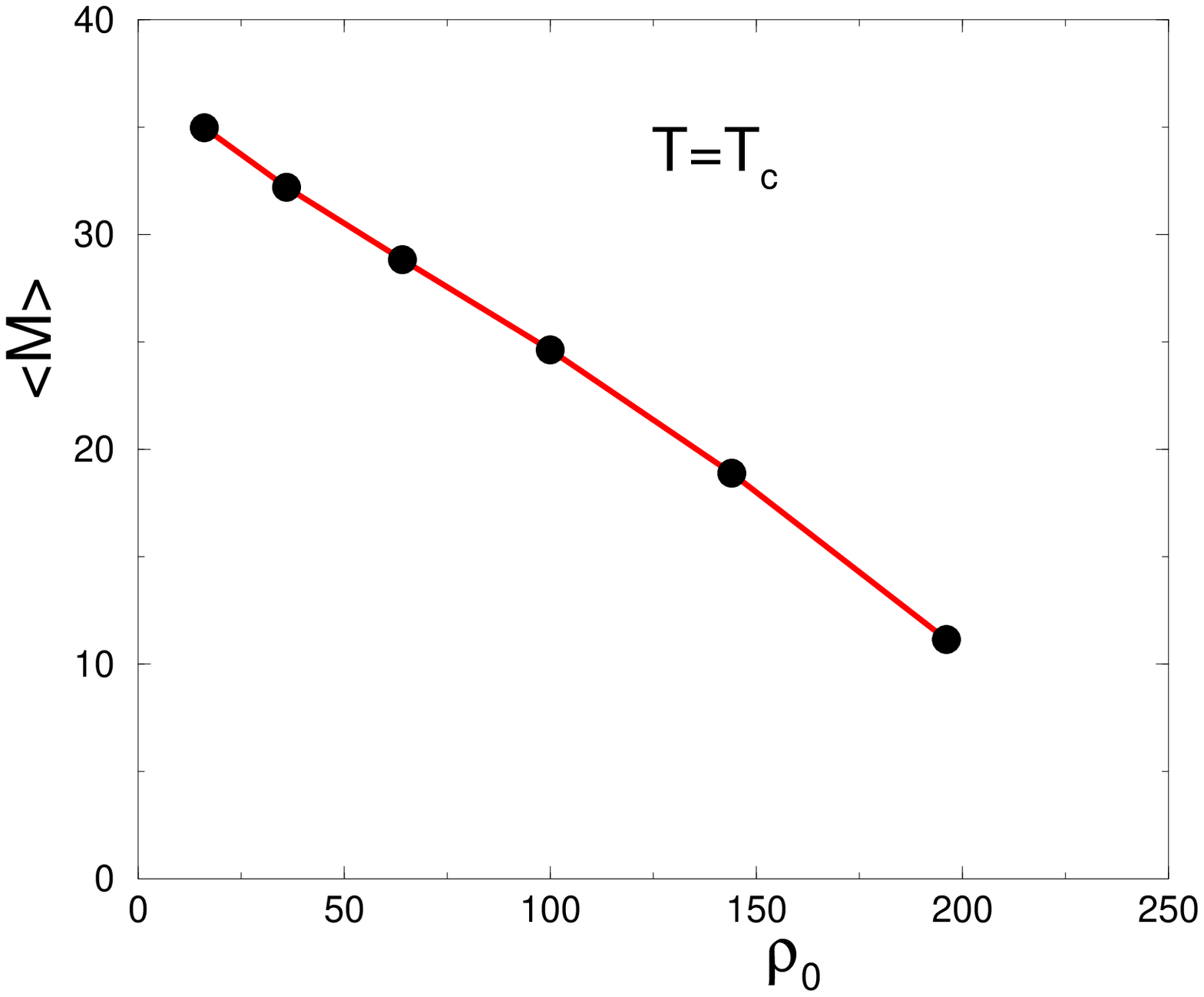}}
\vskip 1.5 cm
\centerline{\bf \large ~~~~~~~~~~Fig.2}
\vskip -0.5 cm
\end{figure}

\newpage

\begin{figure}[t]\epsfxsize=14cm \epsfysize=14cm
\centerline{\epsfbox{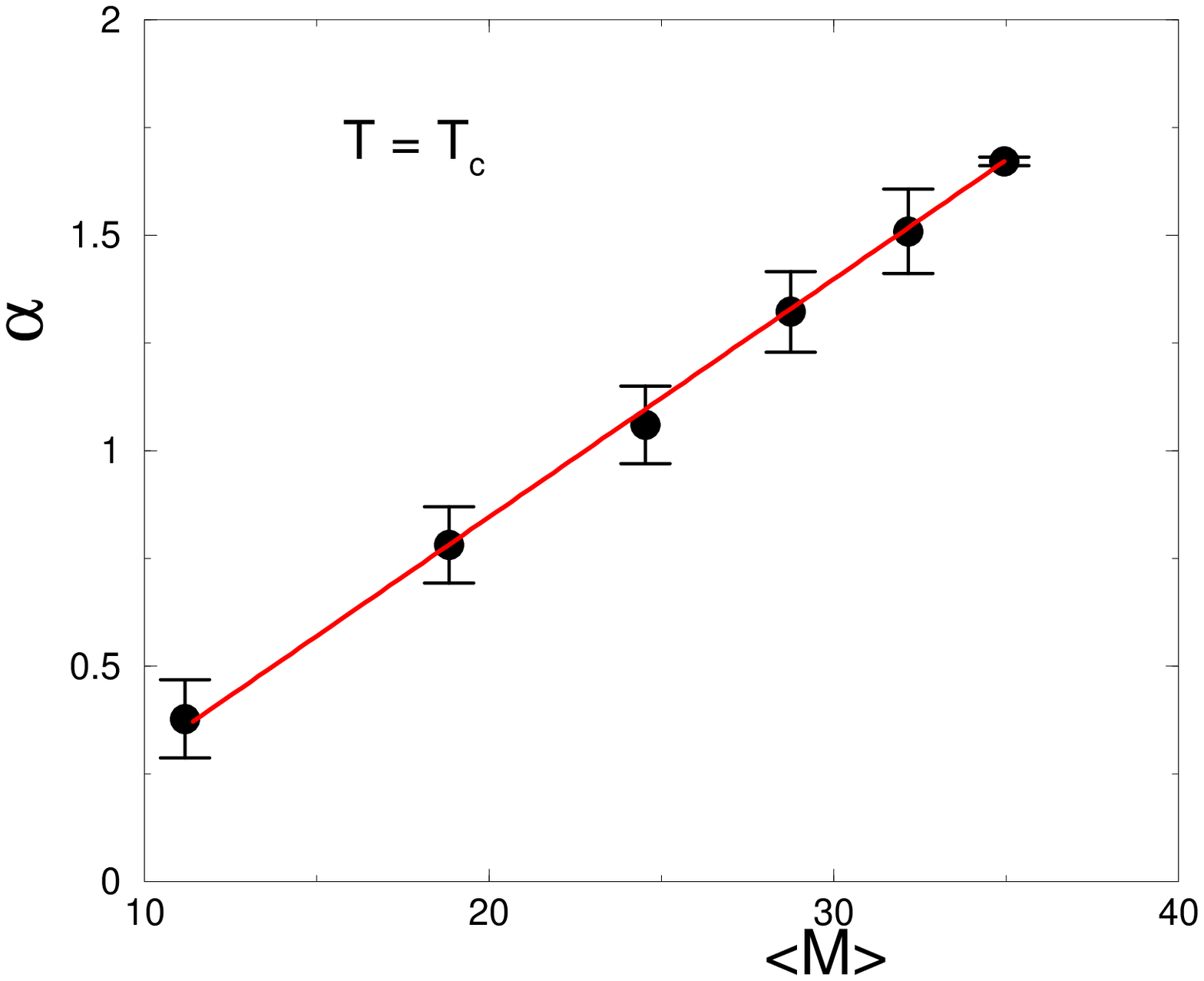}}
\vskip 1.5 cm
\centerline{\bf \large ~~~~~~~~~~~~~~Fig.3}
\vskip -0.5 cm
\end{figure}

\newpage
\begin{figure}[t]\epsfxsize=14cm \epsfysize=14cm
\centerline{\epsfbox{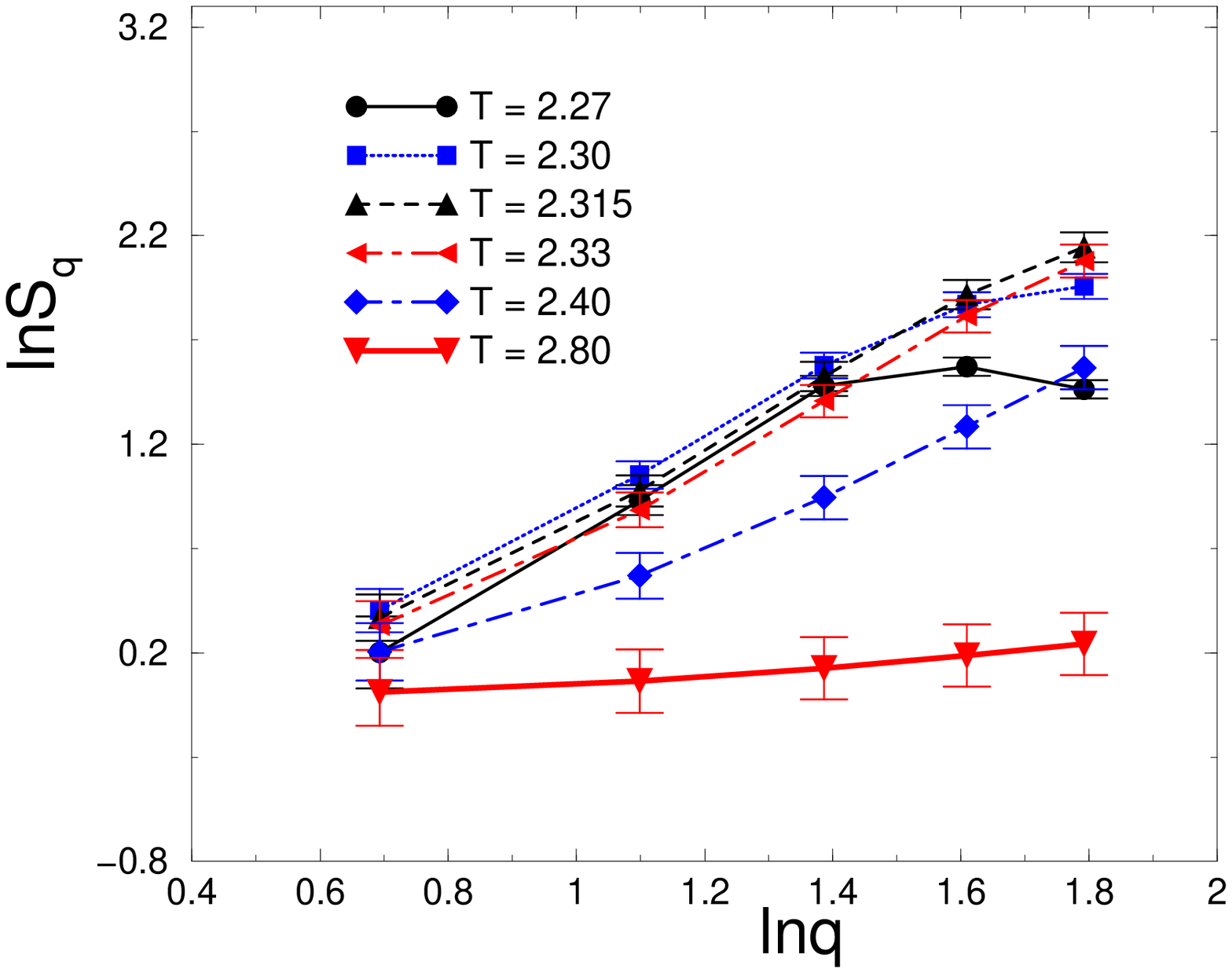}}
\vskip 1.0 cm
\vskip 1.5 cm
\centerline{\bf \large ~~~~~~~~~~~~~Fig.4}
\vskip -0.5cm
\end{figure}

\newpage
\begin{figure}[t]\epsfxsize=14cm \epsfysize=14cm
\centerline{\epsfbox{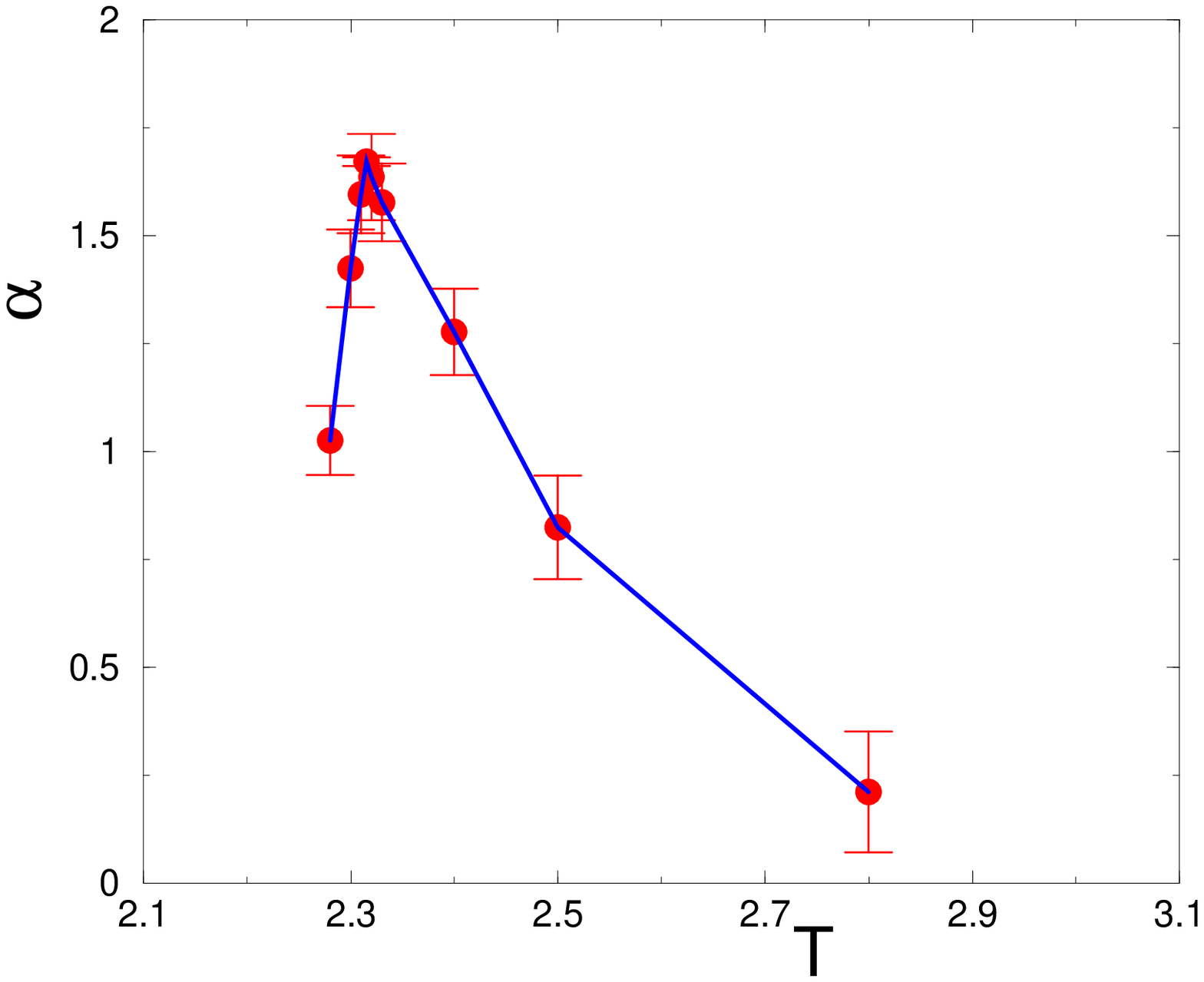}}
\vskip 1.5 cm
\centerline{\bf \large ~~~~~~~~~~~Fig.5}
\vskip -0.5cm
\end{figure}

\newpage
\begin{figure}[t]\epsfxsize=14cm \epsfysize=14cm
\centerline{\epsfbox{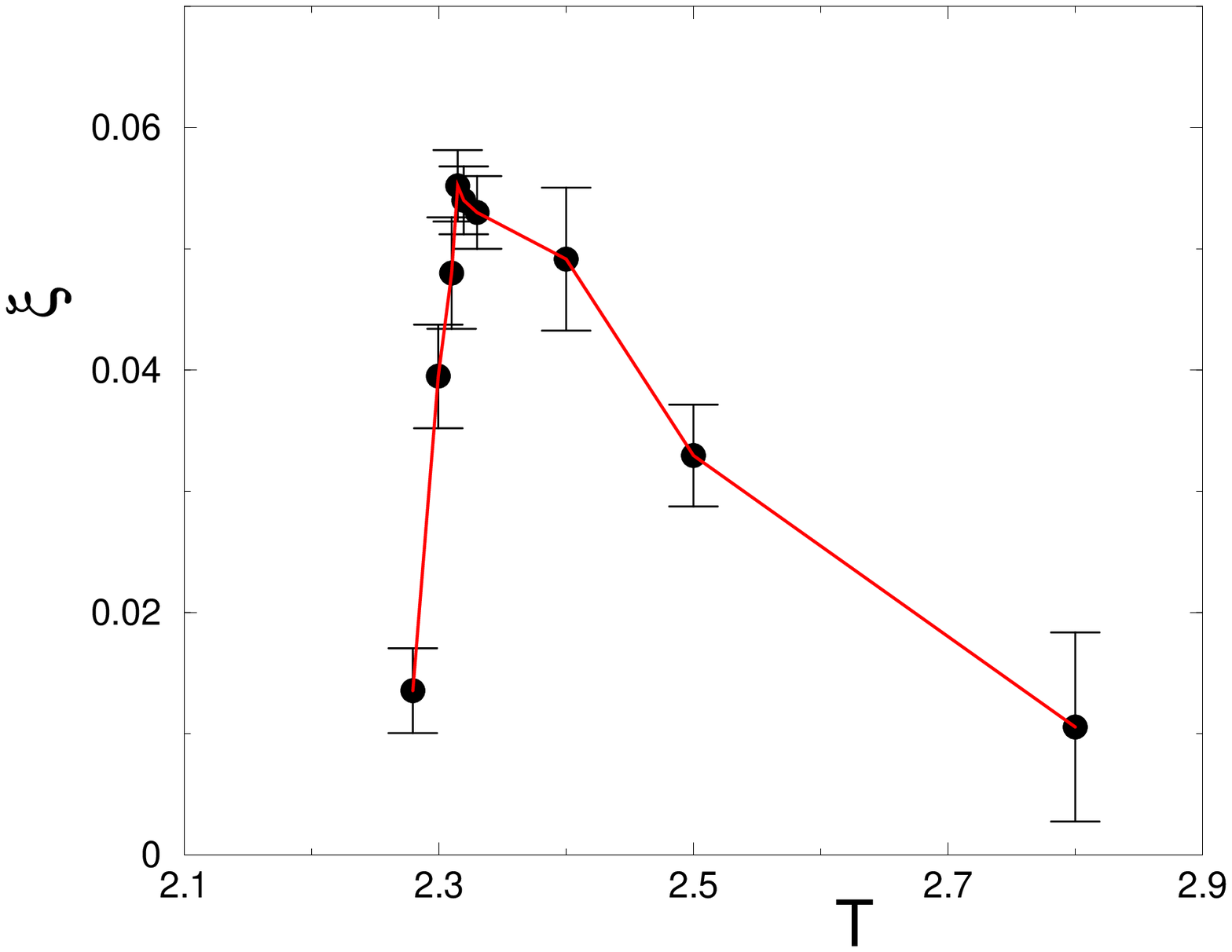}}
\vskip 1.5 cm
\centerline{\bf \large ~~~~~~~~~~~~~Fig.6}
\vskip -0.5cm
\end{figure}

\newpage
\begin{figure}[t]\epsfxsize=14cm \epsfysize=14cm
\centerline{\epsfbox{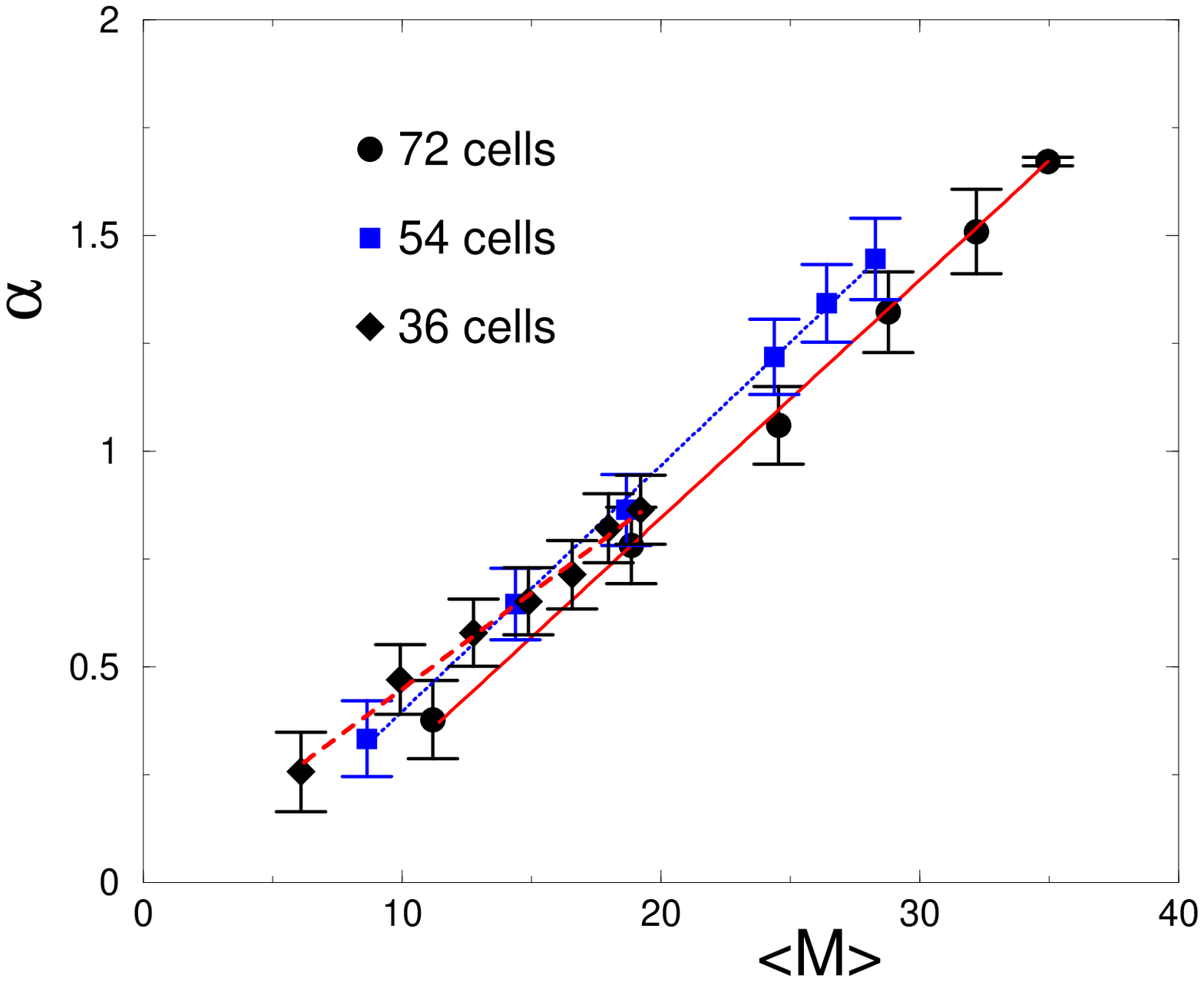}}
\vskip 1.5 cm
\centerline{\bf \large ~~~~~~~~~~~~~Fig.7}
\vskip -0.5cm
\end{figure}

\newpage
\begin{figure}[t]\epsfxsize=14cm \epsfysize=14cm
\centerline{\epsfbox{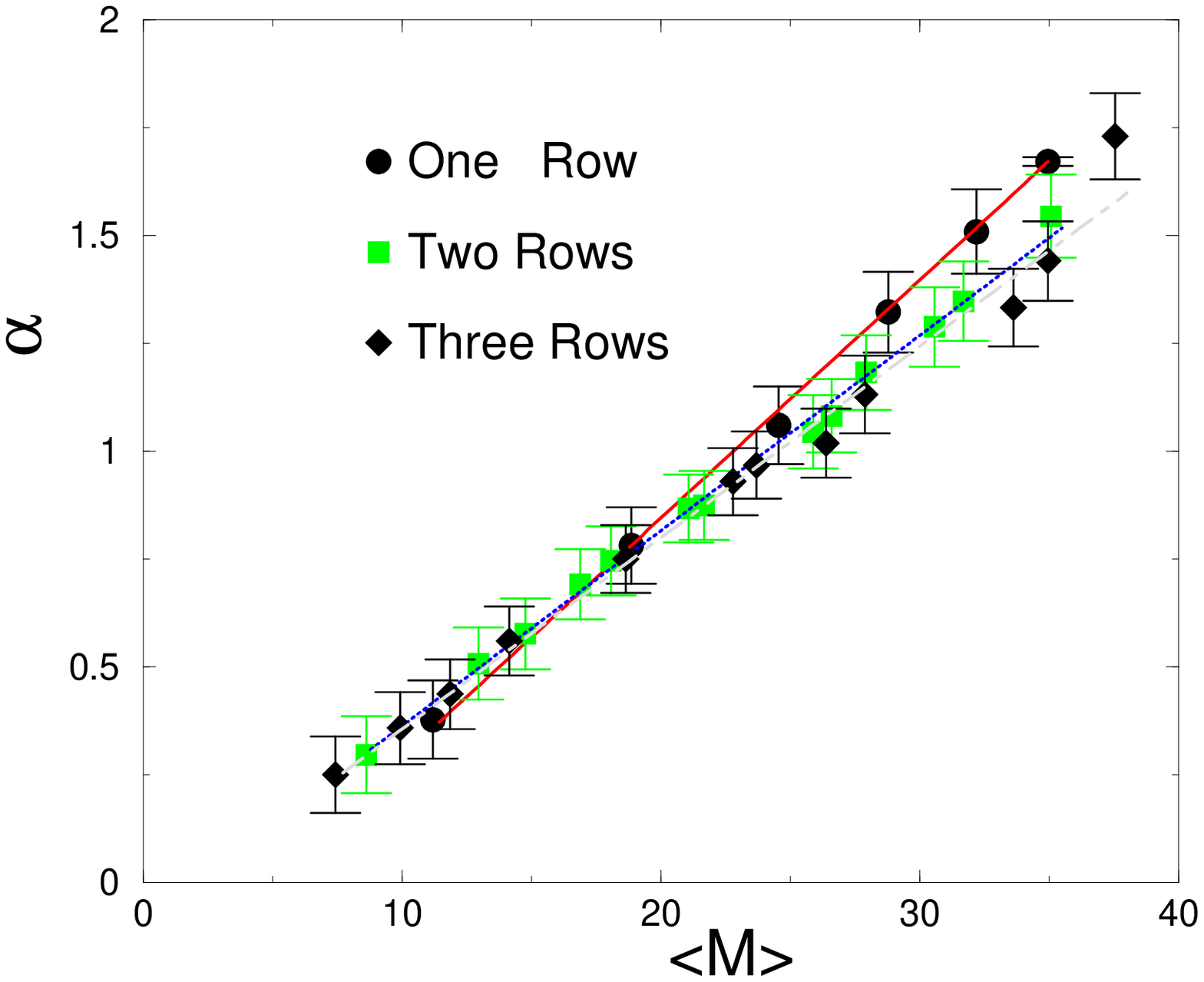}}
\vskip 1.5 cm
\centerline{\bf \large ~~~~~~~~~~~~~Fig.8}
\vskip -0.5cm
\end{figure}

\end{document}